\documentclass[review]{elsarticle}
\usepackage{amsmath, mathrsfs, amssymb,amsfonts,amsthm,graphicx, epsf, dcolumn}
\usepackage[hyperfootnotes=true]{hyperref}
\usepackage{subfigure}
\usepackage{color}
\usepackage{slashed}
\usepackage{setspace}
\usepackage{cancel}
\usepackage{wasysym}
\usepackage{hyperref}
\usepackage{float}
\usepackage[normalem]{ulem}

\pdfoutput=1
\newcommand{\be}{\begin{equation}}
\newcommand{\ee}{\end{equation}}
\newcommand{\bea}{\begin{eqnarray}}
\newcommand{\eea}{\end{eqnarray}}

\begin{document}

\begin{frontmatter}

\title{Towards geometric inflation: the cubic case}

\author{Gustavo Arciniega$^\dag$, Jos\'e D. Edelstein$^\dag$ and Luisa G. Jaime$^\star$}
\address{$^\dag$Dep. F\'\i sica de Part\'\i culas $\&$ Instituto Galego de F\'\i sica de Altas Enerx\'\i as (IGFAE)\\Universidade de Santiago de Compostela, E-15782 Santiago de Compostela, Spain\\ \vskip2mm
$^\star$Instituto de Ciencias Nucleares, Universidad Nacional Aut\'onoma de M\'exico\\ A.P. 70-543 CDMX 04510, Mexico}

\begin{abstract}
We present an up to cubic curvature correction to General Relativity with the following features: (i) its vacuum spectrum solely consists of a graviton and is ghost-free, (ii) it possesses well-behaved black hole solutions which coincide with those of Einsteinian cubic gravity, (iii) its cosmology is well-posed as an initial value problem and, most importantly, (iv) it entails a geometric mechanism triggering an inflationary period in the early universe (driven by radiation) with a {\it graceful exit} to a late-time cosmology arbitrarily close to $\Lambda$CDM.
\end{abstract}


\end{frontmatter}

\section{Introduction}
\label{secc:Introduction}

It is a well-known fact that higher curvature corrections tend to improve the renormalization properties of General Relativity (GR) \cite{Stelle}. It is also widely understood that GR should be seen as providing a low energy effective dynamics of the gravitational field. Therefore, higher curvature corrections are expected to be there and become relevant at somewhat higher energies. Depending upon the scale at which they become relevant, causality might be at stake pointing towards the existence of a richer structure involving higher-spin fields \cite{CEMZ}.

Several modifications of gravity have been proposed throughout the years in order to provide a proper cosmological evolution (see \citep{Clifton:2011jh, Nojiri:2017ncd} and references therein). Particularly important for all these alternatives is the explanation of the accelerated periods in our universe. Some of the proposals were successful explaining late-time acceleration while others accounted for an inflationary period in the early universe. Currently, none of the proposals can both explain inflation adequately and evolve smoothly into a late time acceleration era without invoking extra degrees of freedom.

In this letter, we present an up to cubic curvature correction to General Relativity whose vacuum spectrum solely consists of a graviton and is ghost-free. While it possesses a black hole solution coinciding with that of so-called Einsteinian Cubic Gravity \cite{BC-ECG, HennigarMann, BC-4dBH}, its cosmology is significantly different. Contrary to what turns out to be a generic feature of higher curvature gravities, the set of Friedmann equations remain second order, thereby defining a well-posed initial value problem in a FLRW universe. Most interestingly, it entails a geometric mechanism triggering an inflationary period in the early universe with a {\it graceful exit} to a late-time cosmology arbitrarily close to $\Lambda$CDM.

\section{The theory}
\label{TheTheory}

Restricted to four space-time dimensions, it was recently shown \cite{HKM} that there are four possible cubic Lagrangians, which can be constructed as linear combinations of the ten possible terms,
\begin{equation*}
\begin{array}{ll}
\mathcal{L}_1 = {{{R_a}^c}_b}^d {{{R_c}^e}_d}^f {{{R_e}^a}_f}^b ~, \qquad & \mathcal{L}_2 = {R_{ab}}^{cd} {R_{cd}}^{ef} {R_{ef}}^{ab} ~, \\ [0.43em]
\mathcal{L}_3 = R_{abcd} {R^{abc}}_e R^{de} ~, \qquad & \mathcal{L}_4 = R_{abcd} R^{abcd} R ~, \\ [0.43em]
\mathcal{L}_5 = R_{abcd} R^{ac} R^{bd} ~, \qquad & \mathcal{L}_6 = {R_a}^b {R_b}^c {R_c}^a ~, \\ [0.43em]
\mathcal{L}_7 = R_{ab} R^{ab} R ~, \qquad & \mathcal{L}_8 = R^3 ~, \\ [0.43em]
\mathcal{L}_9 = \nabla_a R_{bc} \nabla^a R^{bc} ~, \quad & \mathcal{L}_{10} = \nabla_a R\, \nabla^a R ~,
\end{array}
\end{equation*}
whose static spherically symmetric vacuum solutions are of the type $ds^2 = -f(r) dt^2 + dr^2/f(r) + \ldots$ and solve a single field equation ${\mathcal{E}^t}_t = {\mathcal{E}^r}_r = 0$ ---at the quadratic level, the same conditions uniquely lead to the Lanczos-Gauss-Bonnet combination, which is a boundary term---. Two of them are actually zero, corresponding to the six-dimensional Euler density, $\chi_6$, and
$$
\mathcal{L}_3 - \frac14 (\mathcal{L}_4 + \mathcal{L}_8) - 2 (\mathcal{L}_5 + \mathcal{L}_6 - \mathcal{L}_7) = 0 ~.
$$
The remaining two can be chosen to be
\begin{eqnarray}
\mathcal{P} & = & 12 \mathcal{L}_1 + \mathcal{L}_2 -12 \mathcal{L}_5 + 8 \mathcal{L}_6 ~, \label{LagP} \\ [0.4em]
\mathcal{C} & = & \mathcal{L}_3 - \frac14 \mathcal{L}_4 - 2 \mathcal{L}_5 + \frac12 \mathcal{L}_7 \label{LagC} ~.
\end{eqnarray}
Their dynamics in vacuum is free of ghosts and any kind of massive modes. It solely consists of a graviton  \cite{BC-ECG}. For this reason, the Lagrangian $\mathcal{P}$ ---discovered earlier--- has been coined Einsteinian Cubic Gravity (ECG)  \cite{BC-ECG}. The inclusion of $\mathcal{C}$, although identified in \cite{HKM}, has not been considered so far in the literature for a reason: its equations of motion are identically null when evaluated on a black hole ansatz. This led to the misleading conclusion that $\mathcal{C}$ is somehow trivial and can be neglected.

The situation changes drastically, in fact, when one considers a cosmological scenario. If we consider a FLRW ansatz in the realm of ECG,
\begin{equation}
ds^2 = -dt^2 + a(t)^2 \left( \frac{dr^2}{1-k r} + r^2 d\Omega^2 \right) ~,
\label{FRWmetric}
\end{equation}
where $d\Omega^2$ is the metric of the unit round sphere, $d\Omega^2 = d\theta^2 + \sin^2\theta d\phi^2$, the resulting equations of motion for $a(t)$ are fourth-order. This may become problematic from the point of view of both dealing with a well-posed initial value problem and experiencing a causal evolution. On the other hand, this is a standard issue in the realm of $f(R)$ gravities \citep{Jaime:2010kn}, where it can be dealt with by converting it into an equivalent scalar-tensor theory; the fields doubling drives the dynamics into a system of second-order ordinary differential equations \citep{Sotiriou:2008rp}.

It is natural to ask whether the Lagrangian $\mathcal{C}$ can play a significant role to alleviate this problem ---although trivial when evaluated in a black hole ansatz, by its own it also leads to  fourth-order differential equations for $a(t)$. Interestingly enough, the answer turns out to be yes! There is a single combination of $\mathcal{P}$ and $\mathcal{C}$ that ---while preserving all the nice features of ECG--- leads to a cosmological scenario where the set of Friedmann equations remains second order; we call the resulting theory {\it Cosmological Einsteinian Cubic Gravity} (CECG),
\begin{equation}
S = \int d^4x \sqrt{-g} \left[ \frac{1}{2\kappa} (R-2\Lambda) + \beta\,(\mathcal{P} - 8\,\mathcal{C} )\right] ~,
\label{actionFRW}
\end{equation}
which is characterized by a single parameter, $\beta$. Now, the propagating physical modes and the effective gravitational coupling $\kappa_{\text{eff}}$ of such a higher curvature gravity can be obtained from the fast linearization procedure given in \cite{Pablo:2017}, where it is shown that, besides the graviton, there may exist two massive modes, a ghosty graviton with mass $m_g$ and a scalar mode with mass $m_s$. The values for $m_g$, $m_s$ and $\kappa_{\text{eff}}$ are model dependent. In the case of (\ref{actionFRW}), the massive modes decouple, $m_g, m_s \rightarrow \infty$, while
$$
\frac{1}{\kappa_{\text{eff}}} = \frac{1}{\kappa} + 48\beta\Lambda^2 ~;
$$
see Table 2 and eqs. ($2.26$) and ($2.27$) in \cite{Pablo:2017}. These are the conditions for the theory to be Einsteinian; {\it i.e.}, free of massive and ghost-like propagating modes. Furthermore, among all cubic curvature theories constructed from $\mathcal{P}$ and $\mathcal{C}$, that governed by action (\ref{actionFRW}) is the only one leading to a second-order ordinary differential equation for $a(t)$ in a FLRW universe.

\section{Cubic cosmology}
\label{TheTheory-Cosmology}

Let us thoroughly analyze the resulting cosmology. The field equations resulting from action (\ref{actionFRW}) are
\begin{equation}
G_{\mu\nu}+E^{\mathcal{P}}_{\mu\nu}+E^{\mathcal{C}}_{\mu\nu}=\kappa T_{\mu\nu} ,
\label{actionvar}
\end{equation}
where $G_{\mu\nu}$ is the Einstein tensor (including the cosmological term), while $E^{\mathcal{P}}_{\mu\nu}$ and $E^{\mathcal{C}}_{\mu\nu}$ result from the variation of $\mathcal{P}$ and $\mathcal{C}$. 
After considering $k=0$ for a flat spatial curvature in the metric (\ref{FRWmetric}), and plugin it into the equations of motion, we obtain a modified Friedmann dynamics:
\begin{equation}
\label{eq:H}
H^2 \left( 1 + 16\beta\kappa H^4 \right) = \frac{1}{3}(\Lambda+\kappa \rho)
\end{equation}
and
\begin{equation}
\label{eq:Hdot}
\dot{H}=\frac{\Lambda-\kappa P-3H^2-48 \beta \kappa H^6}{2+96 \beta \kappa H^4} ~,
\end{equation}
where $H\equiv \dot{a}/a$ and our matter source is an ideal fluid; in natural units:
\begin{equation}
T_{\mu\nu}=\rho  u_{\mu}u_{\nu}+P\left(g_{\mu\nu}+u_{\mu}u_{\nu}\right) ~.
\end{equation}
From (\ref{eq:H}) and (\ref{eq:Hdot}) we can recover the standard Friedmann equations by setting $\beta = 0$. The Hamiltonian constraint (\ref{eq:H}) is used to verify the numerical solution of equation (\ref{eq:Hdot}), which is a first order ordinary differential equation that can be written as $\dot{H}=F(H,\rho,P)$. The initial conditions can be set to $H(a_{in})$, $P(a_{in})$ and $\rho(a_{in})$ for some value $a_{in}\neq 0$ (just like in GR). Taking $\beta \geqslant 0$ we have no divergencies and the usual theorems for differential equations guarantee a well-posed initial value problem.

It is worth noticing at this point that we are bound to have a different evolution for $H(t)$ than the standard one, without the need of adding any kind of scalar field or effective energy-momentum tensor whatsoever. By substituting (\ref{eq:H}) in (\ref{eq:Hdot}) we obtain
\begin{equation}
\label{eq:acceleration}
\frac{\ddot{a}}{a}=H^2-\frac{\kappa(P+\rho)}{2+96\beta\kappa H^4} ~.
\end{equation}
For the inflationary period and for late time acceleration, $\ddot{a}>0$. In the GR framework, we need to invoke some field such that $P<0$ in order to obtain a positive acceleration at early times \cite{Baumann:2009ds}. In the same sense works the addition of the cosmological constant for the late time acceleration: it is necessary in order to prevent the deceleration of the universe.

In the present proposal, the evolution of the acceleration depends on the factor $96\beta\kappa H^4$, which modules the evolution of $\kappa(P+\rho)$ and allows $H^2$ to dominate for some periods, making it possible to obtain inflation and late time acceleration in purely geometric terms. If $\beta$ were negative\footnote{Notice that this is precisely the sign necessary to obtain well-behaved black hole solutions in this theory \cite{HennigarMann, BC-4dBH}. We thank Pablo Bueno for pointing this to us.} we would reach a singularity in the past; $\dot{H}$ would not be determined at that point, the initial value problem would be ill-defined and the Kretschmann scalar would diverge.\footnote{See, for example, the analogous case of Lovelock cosmology in \cite{ThesisCamanho}.} Thereby, we choose $\beta > 0$.

In alternative theories of gravity, it is a widespread habit to define an effective energy-momentum tensor, in such a way that the corrections can be seen as the addition of some kind of fluid. Such fluid should be tested in order to avoid ill behaviors ({\it e.g.}, ghosts or massive modes). In our proposal, this is not necessary: with the simple re-definition of the critical density, accordingly to (\ref{eq:H}),
\begin{equation}
\label{eq:rho-crit}
\rho_c=\frac{3H^2}{\kappa}(1+16\beta\kappa H^4) ~,
\end{equation}
we avoid the inclusion of extra components. In order to explore the capability of this theory to trigger a viable universe and whether it can provide both an inflationary period and late time acceleration, we numerically integrate the field equations. As usual, the evolution of the matter content goes like $\rho_m = \rho_m^0 a^{-3}$ with $P_m=0$ and radiation goes as $\rho_r = \rho_r^0 a^{-4}$ with $P_r = 1/3 \rho_r$. From eq. (\ref{eq:H}) it can be seen already that $a(t)\propto t^{3/2}$ at early times if $\beta > 0$.

\section{Cosmological exploration}
\label{Cosmological exploration}

In the present letter we want to explore robust features of our proposal leaving for the future a more detailed exploration of the full parameter space. Therefore, we will consider and compare just three cases that give us enough pieces of evidence to reach generic and interesting conclusions: A. General Relativity ({\it i.e.}, $\beta=0$), B. CECG with $\beta = 10^{-16}$, and C. CECG with $\beta = 10^{-17}$. In all cases we consider the $\Lambda$CDM framework and the standard values for $\Omega_m^0= 0.3089$, $\Omega_r^0= 8.4\times 10^{-5}$ and $\Omega_\Lambda^0 = 0.691$ found by Planck 2015 \cite{Ade:2015xua}.

We performed a numerical integration of the field equations (\ref{eq:Hdot}) and analyzed the following observables: the age of the universe, its early and late time acceleration, the evolution of the matter, radiation and dark energy densities, and the evolution of the Hubble parameter for low values of the redshift $z$. The integration is performed from the present to the past, the values for $\beta$ were chosen just for the sake of comparison against the GR predictions; it must be ultimately obtained from experimental data. The results are plotted and discussed below.

Under the present proposal the age of the universe is affected in an interesting direction when compared to General Relativity. 
\begin{figure}[h]
\center\includegraphics[width=0.75\textwidth]{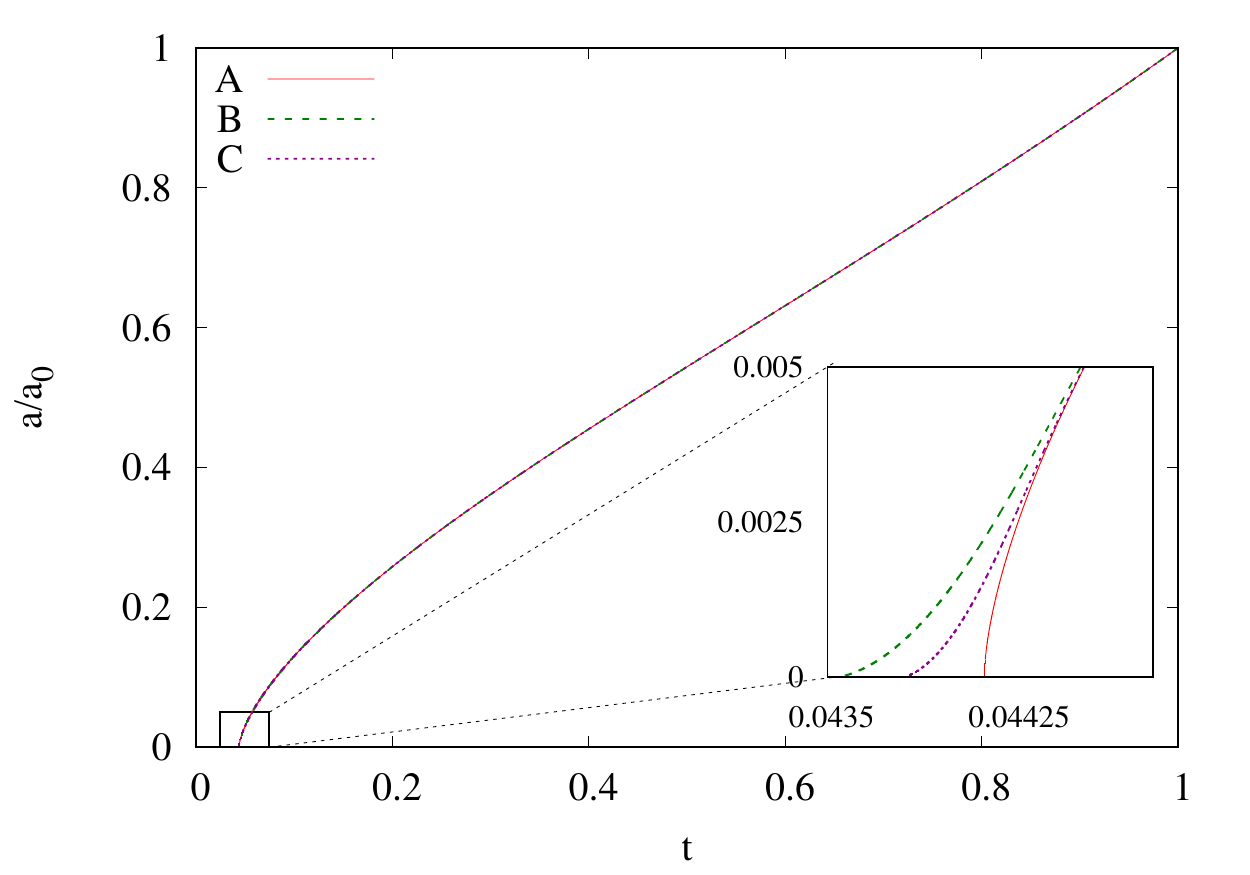}
\caption{\small Comparison of the age of the universe where A is GR, while B and C are CECG with $\beta = 10^{-16}$ and $10^{-17}$, respectively. In all cases we consider the $\Lambda$CDM framework and the currently accepted values for $\Omega_i^0$. The interior frame displays a zoom into early time.}
\label{fig:edades}
\end{figure}
As shown in Figure {\ref{fig:edades}, both in the B and C scenarios, $a(t)$ displays a characteristic shape that generically makes this theory able to produce an accelerated epoch at early times, $a(t) \sim t^{3/2}$. This is confirmed in Figure \ref{fig:Inflation}, where we analyze the acceleration given by (\ref{eq:acceleration}).
\begin{figure}[h]
\center\includegraphics[width=0.85\textwidth]{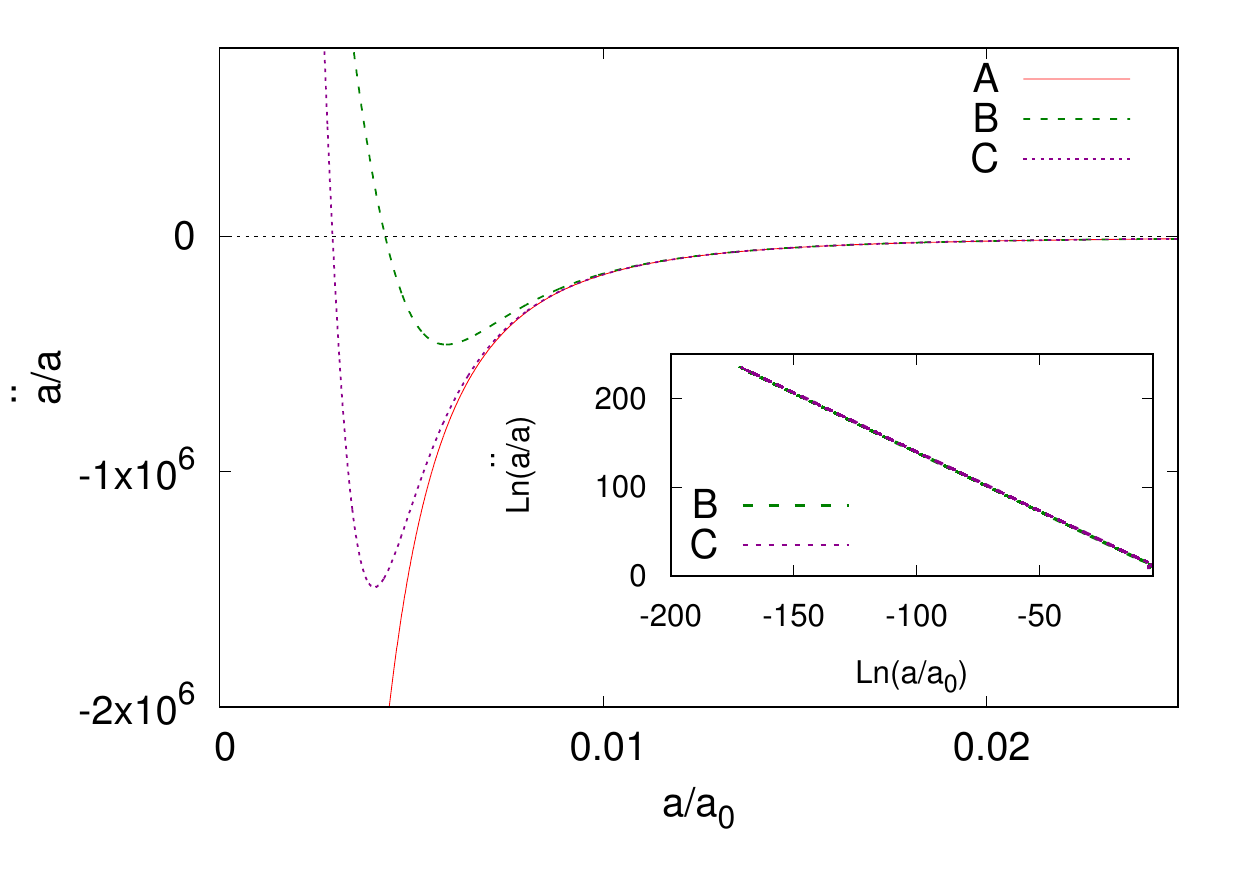}
\caption{\small Big frame: Evolution from early to late time acceleration for the A, B and C cases. Interior frame: Early time acceleration for B and C cases ($log-log$ scale).}
\label{fig:Inflation}
\end{figure}
We display the acceleration at two selected ranges, from $0<a/a_0<0.02$ where we can notice the evolution in comparison to GR. In the cases B and C the acceleration comes from very large positive values, then goes to negative and from there on the evolution fits the one followed in GR. For late-times, $a/a_0>0.02$, the evolution is practically the same as in GR+$\Lambda$CDM; the comparison at these values of $z$ will make more sense when we explore the evolution of the Hubble parameter $H/H_0$ (see Figure \ref{fig:H}).

The interior frame of Figure \ref{fig:Inflation} shows the evolution at early times for the cases B and C. In both scenarios the acceleration goes like $\ddot{a} \sim a^{-1/3}$ during several e-folds. This evolution is limited on physical grounds just by the Planck length, $\textit{l}_P$, otherwise the singularity $a=0$ would apparently be reached. The inflationary period in this theory transits smoothly to the standard behavior in GR at late times, solving in this way the {\it graceful exit} problem.

We show the evolution of the densities $\Omega_m$, $\Omega_r$ and $\Omega_\Lambda$ for two cases in Figure \ref{fig:Ome-GR-A2}, the one we called B and, in order to trigger the discussion, an extra case D with $\Omega_\Lambda =0$, $\beta = 10^{-3}$, $\Omega_m^0= 0.9$ and $\Omega_r^0=0.1$.
\begin{figure}[h]
\center\includegraphics[width=0.85\textwidth]{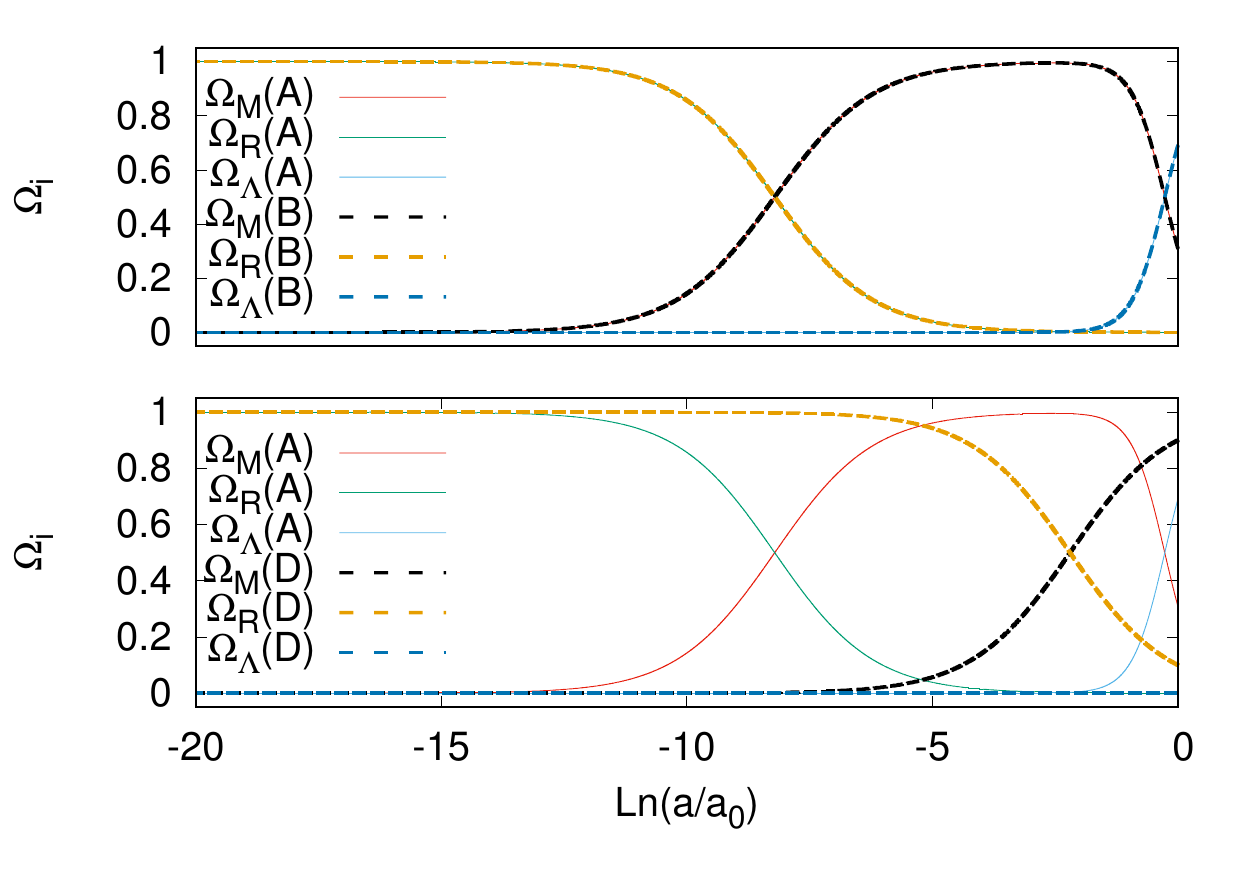}
\caption{\small Evolution of the densities $\Omega_m$, $\Omega_r$ and $\Omega_\Lambda$ for the B (top pannel) and D (bottom pannel) cases (both plotted using dashed lines) in comparison to General Relativity, which is displayed in solid-lines.}
\label{fig:Ome-GR-A2}
\end{figure}
This way we are able to notice the importance of the cosmological constant in the present framework. We split them into two panels in order to show the direct comparison of both cases B and D with GR. From the top figure it is clear that the evolution of $\Omega_m$, $\Omega_r$ and $\Omega_\Lambda$ are, basically, the same in A and B. Matter dominated epoch occurs and $\Omega_m = \Omega_r$ is reached at the same value of $z$ (or $\ln(a)$ in the case of the figure). Instead, when we consider the case D, the evolution of $\Omega_m$, $\Omega_r$ and $\Omega_\Lambda$ go differently: even when matter domination is well behaved, the matter-radiation equality time is reached at lower values of $z$. Although this can be seen as a reason to rule out this latter model, a thorough analysis of the parameter space seems necessary before reaching a definitive conclusion.

In order to have the insight to compare our predictions, at the background level, with observational data for late time cosmology, we compute $H(z)/H_0$. Figure \ref{fig:H} shows the evolution of the Hubble parameter for the cases A, B and C.
\begin{figure}[h]
\center\includegraphics[width=0.85\textwidth]{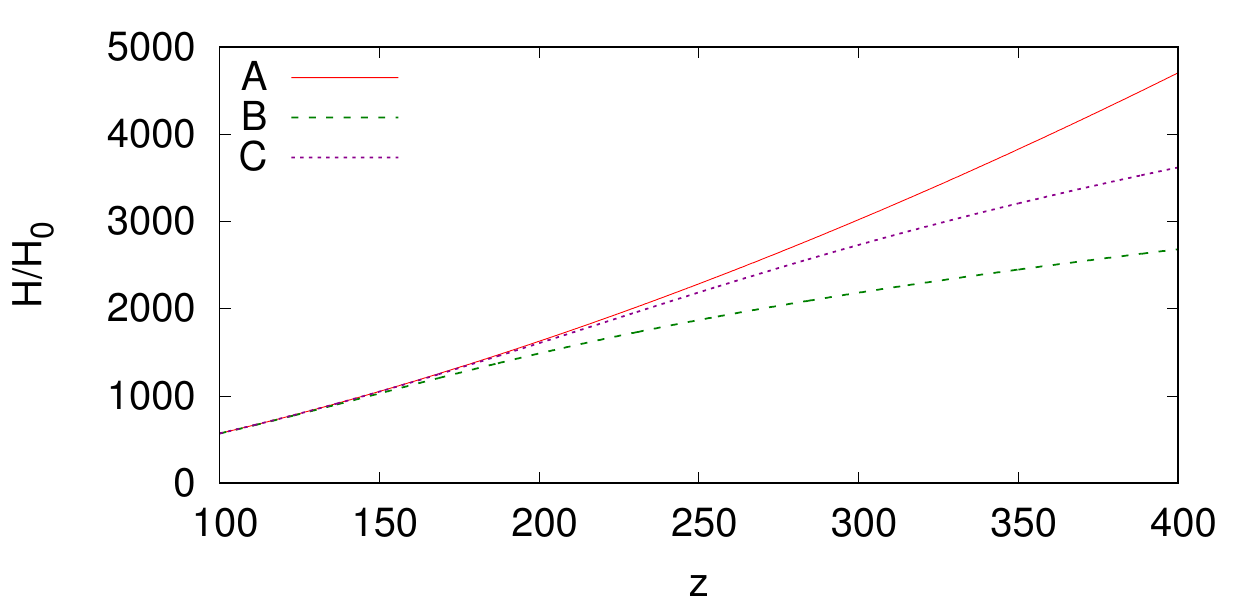}
\caption{\small Evolution of the Hubble parameter $H/H_0$ for A, B and C cases.}
\label{fig:H}
\end{figure}
As expected, they are arbitrarily close at late times, thereby making possible to match for instance with supernovae Ia observations \cite{Riess, Perlmutter}. For higher values of $z$ (in the figure, $z \sim 150$, but this value certainly depends on the choice of $\beta$), the evolution in cases B and C goes slower than the one in GR. Close to the CMB the values of $H(z)/H_0$ for the cases B and C are very different from those obtained in General Relativity; this is simply telling us that a realistic\footnote{The world {\it realistic} should be taken with a grain of salt; we introduce a novel mechanism of geometric inflation that for large curvatures must certainly be complemented by the introduction of the whole series of higher-curvature corrections \cite{Arciniega:2018tnn} and possibly by the inclusion of a scalar field \cite{AlejandroVL}.} $\beta$ must be even smaller than $10^{-17}$.

\section{Discussion}
\label{Discussion}

In this letter we presented a novel gravitational theory which, while enjoying all the nice features of the recently proposed Einsteinian Cubic Gravity \cite{BC-ECG}, possesses an additional remarkable property concerning its cosmology: the scale factor obeys a second-order field equation in the case of an isotropic and homogeneous four-dimensional universe. The modification depends on a single parameter; it is purely geometric and has the potential to provide both an accelerated inflationary epoch at early times as well as a late time evolution that is arbitrarily close to GR in the $\Lambda$CDM framework. For these reasons we called it {\it Cosmological Einsteinian Cubic Gravity}. Both the early and late time evolutions are phenomenologically viable provided the parameter $\beta$ is sufficiently small. In the past, the universe appears to have displayed accelerated expansion, $a(t) \sim t^{3/2}$; albeit a power law, a number of e-folds can be obtained from the instant when the scale factor was given by the Planck length, $\textit{l}_P$. The inflationary period has a {\it graceful exit}.

We have also briefly discussed a case in which the cosmological term vanishes in order to show that, despite of the expected difficulties to deal with the late time cosmology, the matter-radiation equality time is reached at lower values of $z$ than those customarily expected (although these results depend on the choice of parameters ---including the matter, radiation and dark energy densities--- and a full analysis is required before arriving at categorical conclusions).

In comparison with the usual modifications of gravity proposed to solve cosmological controversies, the value of $\beta$ in this theory is not just a parameter to be fixed by observations; in our case, it can be seen as a new fundamental gravitational constant. It is natural to expect that it originates from a UV complete theory of gravity as the result of an effective Wilsonian low-energy integration. Further analysis should be performed in order to constrain $\beta$. Current observations of late time cosmological tracers such as supernovae \citep{Suzuki:2011hu}, cosmic chronometers \citep{Moresco:2016mzx} and baryonic acoustic oscillations \citep{Alam:2016hwk, Delubac:2014aqe}, as well as the Cosmic Microwave Background \citep{Ade:2015xua}, can be used to explore the best-fitted values of the parameters $\Omega_m^0$, $\Omega_r^0$, $\Omega_\Lambda$ and $\beta$.

It is worth mentioning that the evolution of $H(z)$ predicted in this theory might help in reducing the conflicting difference of $\sim 3.5 \sigma$ between the value of the Hubble constant, $H_0$, predicted by Planck 2015 \cite{Ade:2015xua} and the one found when more local sources are analyzed (see, for example, \citep{Delubac:2014aqe}). On the other hand, a forecast could be cooked up in order to have ready-to-test predictions in the light of the releases of EUCLID \citep{Laureijs:2011gra} or DESI \citep{Aghamousa:2016zmz} experiments.

Finally, we must mention that different scenarios should be explored in order to pass other gravitational tests \cite{Will:2014kxa}, and further constrain the value of $\beta$ in an independent way. For instance, the existence of compact objects has been an issue in some other modifications of gravity \citep{Jaime:2010kn}, and it is certainly a problem to be addressed in our currently proposed cubic theory. At this stage, our proposal should be understood as a novel mechanism whose phenomenological implementation ultimately needs to be examined in greater detail.

\section*{Acknowledgments}
\noindent
We would like to thank Pablo Bueno, Pablo Cano, Gast\'on Giribet, Marcelo Salgado, David V\'azquez and Alejandro Vilar for fruitful discussions.
GA wishes to acknowledge the hospitality of the IGFAE--USC and the postdoctoral fellowship from CONACYT (grant 207620).
The work of JDE is supported by MINECO FPA2014-52218 and FPA2017-84436-P, Xunta de Galicia ED431C 2017/07, FEDER, and the Mar\'\i a de Maeztu Unit of Excellence MDM-2016-0692.
He wishes to thank the Physics Department of the University of Buenos Aires, where part of this work was done.

\end{document}